\documentclass[reprint,nofootinbib,amsmath,amssymb,aps,prd]{revtex4-2}

\usepackage{graphicx}
\graphicspath{ {./images/} }
\usepackage{dcolumn}
\usepackage{bm}
\usepackage{hyperref}
\usepackage{aas_macros}
\usepackage{amsmath}

\usepackage{xcolor}

\begin{document}

\title{Directed search for continuous gravitational waves from the possible kilonova remnant G4.8+6.2}

\author{Yu Liu}
\email{yul@hust.edu.cn}
\affiliation{Department of Astronomy, School of Physics, Huazhong University of Science and Technology, Wuhan 430074, China}
\affiliation{MOE Key Laboratory of TianQin Project, Sun Yat-sen University, Zhuhai 519082, China}
\author{Yuan-Chuan Zou}%
\email{zouyc@hust.edu.cn}
\affiliation{Department of Astronomy, School of Physics, Huazhong University of Science and Technology, Wuhan 430074, China}
\affiliation{MOE Key Laboratory of TianQin Project, Sun Yat-sen University, Zhuhai 519082, China}

% \date{\today}

\begin{abstract}
G4.8+6.2 was proposed as a possible kilonova remnant associated with the Korean guest star of AD 1163 in our Milky Way galaxy.
Its age is about 860 years according to the historical record. 
If a neutron star was left in the center of G4.8+6.2, this young neutron star may radiate strong continuous gravitational waves, which could beat the indirect age-based upper limit with current LIGO sensitivity.
In this work, we searched such continuous gravitational waves in the frequency band $20-1500 \mathrm{~Hz}$.
This search used two days of LIGO O3b data from the Hanford and Livingston detectors. 
While no signal was found, we placed upper limits on the gravitational wave strain.
For comparison we also showed the latest results of all-sky searches obtained with various search pipelines.
With upgrading of the gravitational wave detectors, it will provide the opportunity to see whether a black hole or a neutron star is harbored inside G4.8+6.2.
\end{abstract}


\maketitle

\section{Introduction}
Since the first detection of gravitational wave (GW) event GW150914 \cite{Abbott2016}, GW signals from compact binary mergers have become routine \cite{GWTC1,GWTC2,GWTC3}.
However, other exciting GW sources still remain undetected.
Continuous gravitational waves (CWs) from rapidly rotating neutron stars (NSs) are one of them \cite{Lasky2015}.
The next milestone event may be this kind of source by improving the detector sensitivities, the search algorithms, and longer observation times \cite{Piccinni2022}.
For GW170817, no postmerger GW signal of a long-lived remnant NS was detected on timescales of days after the binary neutron star (BNS) merger \cite{Abbott2019}. 
If it happens in the Milky Way, it will be an unprecedented opportunity for NS studies \cite{Sieniawska2019}.

Although no signal has been detected, limits have been placed on how deformed the target NS could be.
For instance, LIGO-Virgo-KAGRA Collaboration {\it et al.} \cite{LVK2021} reported new upper limits on the ellipticity of known pulsars.
In addition to the targeted searches for CWs from known pulsars, another type of search strategy is the directed searches for young supernova remnants (SNRs) containing candidate nonpulsing NSs \cite{Riles2017,Sieniawska2019}.
The most promising examples of these sources are Cassiopeia A (Cas A), Vela Jr., and SNR G347.3-0.5 \cite{Abadie2010,SNRSearch,Ming2015}.
A brief summary of the latest search results can be found in Piccinni \cite{Piccinni2022}.

Previous directed searches included a total of 15 SNRs selected from the SNRcat and Green catalogs \cite{SNRO3,Green2019,Ferrand2012}.
G4.8+6.2 was not considered in this gold sample, not only because of the large uncertainty in its age and distance estimates \cite{Bhatnagar2000}, but also because there is no evidence for the existence of a central compact object (CCO) \cite{De2017}.
However, G4.8+6.2 is interesting because it might be a nearby and young kilonova remnant (i.e., remnants of kilonova explosions, KNR) in our Galaxy that contains a massive fast-rotating strongly magnetic NS.
These physical properties make it an ideal directed search target for CWs \cite{Liu2019}.

The paper is organized as follows: 
Sec.~\ref{sec:G4862} presents the source properties of G4.8+6.2 and its putative neutron star.
Section~\ref{sec:Results} places upper limits on the strength of CWs.
Section~\ref{sec:Conclusion} concludes with a discussion of the results and prospects for future searches.

\section{G4.8+6.2}
\label{sec:G4862}

The basic information of the possible KNR G4.8+6.2 is presented in this section.
In Sec.~\ref{sec:KNR}, we briefly describe the physical properties of G4.8+6.2 and the reason why it was suggested as a nearby and young KNR.
Then we discuss the possibility that there is an isolated NS within the center of G4.8+6.2 (Sec.~\ref{sec:INS}), and calculate the power of the pulsar wind nebula (PWN) (Sec.~\ref{sec:PWN}).
We further assess the detectability of CWs from G4.8+6.2 in Sec.~\ref{sec:CWs}.

\subsection{Kilonova remnants}
\label{sec:KNR}

At radio wavelengths, G4.8+6.2 consists of an almost circular shell of $18 \mathrm{~arcmin}$ in angular diameter, centered at $\mathrm{RA} \simeq 17^{\mathrm{h}} 33^{\mathrm{m}} 24^{\mathrm{s}}$, $\mathrm{Dec} \simeq -21^{\circ} 34^{\prime}$ \cite{Bhatnagar2000}.
The brightness distribution in G4.8+6.2 is not uniform (the radio emission is obviously weak on the north and south edges).
It is also a potential very-high energy (VHE) $\gamma$-ray source based on deep observations by the High Energy Stereoscopic System (HESS) with an exposure of 152 hours \cite{Aharonian2022}.
If the $\gamma$-ray emission is dominated by hadronic processes, this may imply that kilonova remnants are the origin of Galactic cosmic rays \cite{Gabici2016}, but for young remnants, it is also likely produced via leptonic processes.
Up to now, no other wavelengths have been observed in this region, which is unlike other well-known young or historical SNRs (SN 1006, Cas A, Tycho, Kepler) with rich multiwavelength observations.

G4.8+6.2 has a relatively high Galactic latitude and lives in a low-density interstellar medium (ISM) environment.
This can be confirmed by its unusual polarization properties \cite{Zhang2003} and barrel-shape morphology \cite{Kesteven1987}.
It is also consistent with the predicted spatial distribution of KNRs \cite{Wu2019}.
However, to unequivocally identify the remnant class, we still need to detect the decay lines from long-lived r-process nuclei in the hard x-ray or $\gamma$-ray bands \cite{Burns2020}.
Korobkin {\it et al.} \cite{Korobkin2020} calculated and found that next-generation $\gamma$-ray telescopes (COSI, AMEGO, LOX) will be able to observe these objects up to $10 \mathrm{~kpc}$.
Recently, Terada {\it et al.} \cite{Terada2022} also gave an estimation of the detectable distance limit of G4.8+6.2.

The connection between G4.8+6.2 and the guest star AD 1163 was reported by Liu {\it et al.} \cite{Liu2019}.
AD 1163 was recorded in a rather unique way concerning the occultation of a guest star by the Moon on a specific date.
Although only a single brief Korean record is extant, we can still deduce its position, brightness and duration of visibility \cite{Stephenson1971}, which suggests a kilonova nature. 

If AD 1163 and G4.8+6.2 are associated, then the remnant age is $\sim 860$ years.
According to the result of Korobkin {\it et al.} \cite{Korobkin2020} and assuming typical values $M_{\mathrm{ej}}=0.01 M_{\odot}$, $v=0.1 \mathrm{c}$, $\rho_{\mathrm{ISM}}=10^{-4} \mathrm{cm}^{-3}$, G4.8+6.2 should be in the Sedov-Taylor phase and imply a remnant diameter of $42.6 \mathrm{~pc}$.
With an angular diameter of $18 \mathrm{~arcmin}$, the distance is estimated to be $\sim 8.1 \mathrm{~kpc}$.
Due to the large uncertainties in parameter, this is only used as a reference value in this paper.
% Of course, a more reliable distance estimation is required for better constraining its basic parameters.

\subsection{Isolated neutron star}
\label{sec:INS}

A non-negligible fraction of BNS mergers go on to form a long-lived NS remnant \cite{Gao2016}.
Observational evidence comes from those short gamma-ray bursts (SGRBs) with the x-ray afterglow plateau in light curves that have been suggested as a millisecond magnetar central engine \cite{Dai2006,Rowlinson2013}.
Different from supernova (SN) explosions, BNS mergers are in favor of forming an isolated NS with high masses, high spins, and strong magnetic fields \cite{Lu2015,Ho2018}.

It is difficult to classify NSs formed by BNS mergers from the current pulsar sample.
For old NSs, they will undergo spin down and magnetic field decay.
Meanwhile, the most precise measurements of NS masses rely on it being in a binary system \cite{Thorsett1999,Lattimer2010}.
Therefore, the most efficient avenue is the discovery of a NS in KNR.

G4.8+6.2 as a KNR candidate, neither pulsars nor their associated wind nebulae have been detected yet \cite{ATNF}.
The nondetection could be due to the beaming and propagation effects that often occur in SNRs \cite{Gaensler1995,Straal2019}.
The other possibility is a black hole (BH) being the merger remnant. 
It is worthwhile to develop a method to distinguish these two central objects.

It is worth noting that nearby Kepler remnant is a Type Ia supernova \cite{Patnaude2012}, which would leave no NS behind.
Hence, CCO found in this area may be association with G4.8+6.2.
Unfortunately, x-ray emission from G4.8+6.2 was strongly contaminated by Kepler \cite{Liu2019}.

\subsection{Pulsar wind nebula}
\label{sec:PWN}

\begin{figure}[h]
    \centering
    \includegraphics[width=0.5\textwidth]{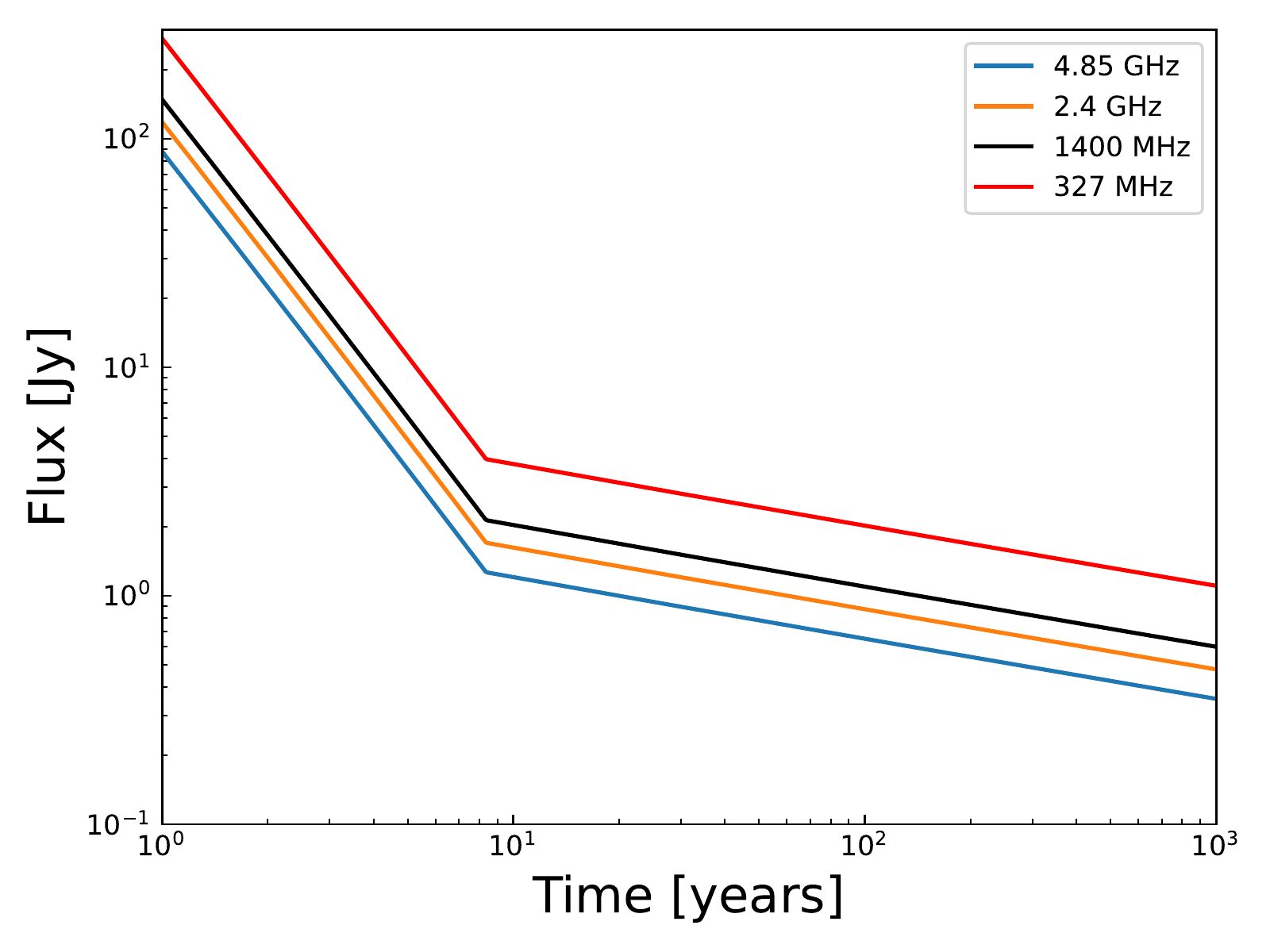}
    \caption{\label{fig:PWN} Multiband radio light curves emitted from the kilonova ejecta-PWN system of the GW170817 event remnant introduced by Ren and Dai \cite{Ren2022} rescaled to $8.1 \mathrm{~kpc}$.}
\end{figure}

Ren {\it et al.} \cite{Ren2019} suggested the possibility of the existence of a PWN embedded in the kilonova ejecta (i.e., a kilonova ejecta-pulsar wind nebula system).
Subsequently Ren and Dai \cite{Ren2022} obtained the optimal parameter values of GRB 170817A by fitting the multi-band light curves of AT 2017gfo using the kilonova ejecta-PWN model, and calculated the late-time radio emission.
We extended their radio emission results to 860 years using the same parameters rescaled to $8.1 \mathrm{~kpc}$ as shown in Fig.~\ref{fig:PWN}.

The break in the radio light curve is the transition from fast-cooling regime to slow-cooling regime (see the evolution of the characteristic synchrotron frequency $\nu_{b}$ and the synchrotron cooling frequency $\nu_{c}$ in Fig.3 of Ren and Dai \cite{Ren2022}).
The selection of frequencies is based on the observations of Bhatnagar \cite{Bhatnagar2000}.
Figure~\ref{fig:PWN} shows that the radio emission of PWN can be barely observed by the NRAO VLA Sky Survey (NVSS) and the Giant Metrewave Radio Telescope (GMRT) \cite{Bhatnagar2000}.
Considering that G4.8+6.2 is a different source with GRB 170817A, the optimal parameter values obtained from GRB 170817A may not be applicable to G4.8+6.2, especially if the merger remnant of GRB 170817A is a black hole.

\subsection{Continuous gravitational waves}
\label{sec:CWs}

We assume the existence of a 860-year-old NS in G4.8+6.2.
Although it is electromagnetically invisible, we might be able to detect it through gravitational wave observation.
Following Bejger \cite{Bejger2018}, we will refer this kind of NSs as ``GW pulsars.''

This NS born from BNS merger favors in the vicinity of the NS maximum mass $\sim 2.5 \mathrm{M}_{\odot}$ \cite{Godzieba2021,Rocha2021}.
Its young age means that strong magnetic fields produced at birth are likely still present.
Such strong magnetic fields could induce large deformations and correspondingly large ellipticity \cite{Gualtieri2011}.
At the same time it will undergo rapid spin down due to magnetic dipole radiation and gravitational radiation, but still possible in the LIGO frequency band as the young age \cite{Cutler2002}.
These features make it a promising target for CWs.

Unfortunately, we do not know the spin frequency and its derivative.
Therefore, we use the frequency-independent age-based upper limit.
The maximum expected GW strain for a NS at distance $D$ with characteristic age $\tau$ and a principle moment of inertia $I_{zz}$ is given by \cite{Wette2008}

\begin{equation}
    \label{equ:UL_age}
    \begin{aligned}
        h_{0}^{\text {age}} &\leq \frac{1}{D} \sqrt{\frac{5 G I_{z z}}{8 c^{3} \tau}} \\
        &\approx 3.02 \times 10^{-25}\left(\frac{8.1 \mathrm{~kpc}}{D}\right)\left(\frac{860 \mathrm{~yr}}{\tau}\right)^{1 / 2}\left(\frac{I_{z z}}{10^{45} \mathrm{~g} \mathrm{~cm}^{2}}\right)^{1 / 2},
    \end{aligned}
\end{equation}
where $G$ is Newton's gravitational constant, and $c$ is the speed of light.
Here, we choose $I_{zz}=10^{45} \mathrm{~g} \mathrm{~cm}^{2}$ for typical values, although the true value could be higher by a factor of 2 \cite{Bejger2005}.

To determine whether a search is worthwhile, $h_{0}^{\text {age}}$ must be greater than the expected sensitivity of the detector which is given by
\begin{equation}
    h_{0}^{95 \%}=\Theta \sqrt{\frac{S_{h}(f)}{T_{\mathrm{obs}}}},
\end{equation}
for a $95 \%$ confidence limit, where $S_{h}$ is the noise power spectral density, $T_{\mathrm{obs}}$ is the coherently integrated observation time, and $\Theta$ is a detection criterion, which depends on the data analysis pipeline. 
For a directed search like ours, $\Theta$ is approximately $30$ \cite{Wette2008,Owen2022}.

\begin{figure}[h]
    \centering
    \includegraphics[width=0.5\textwidth]{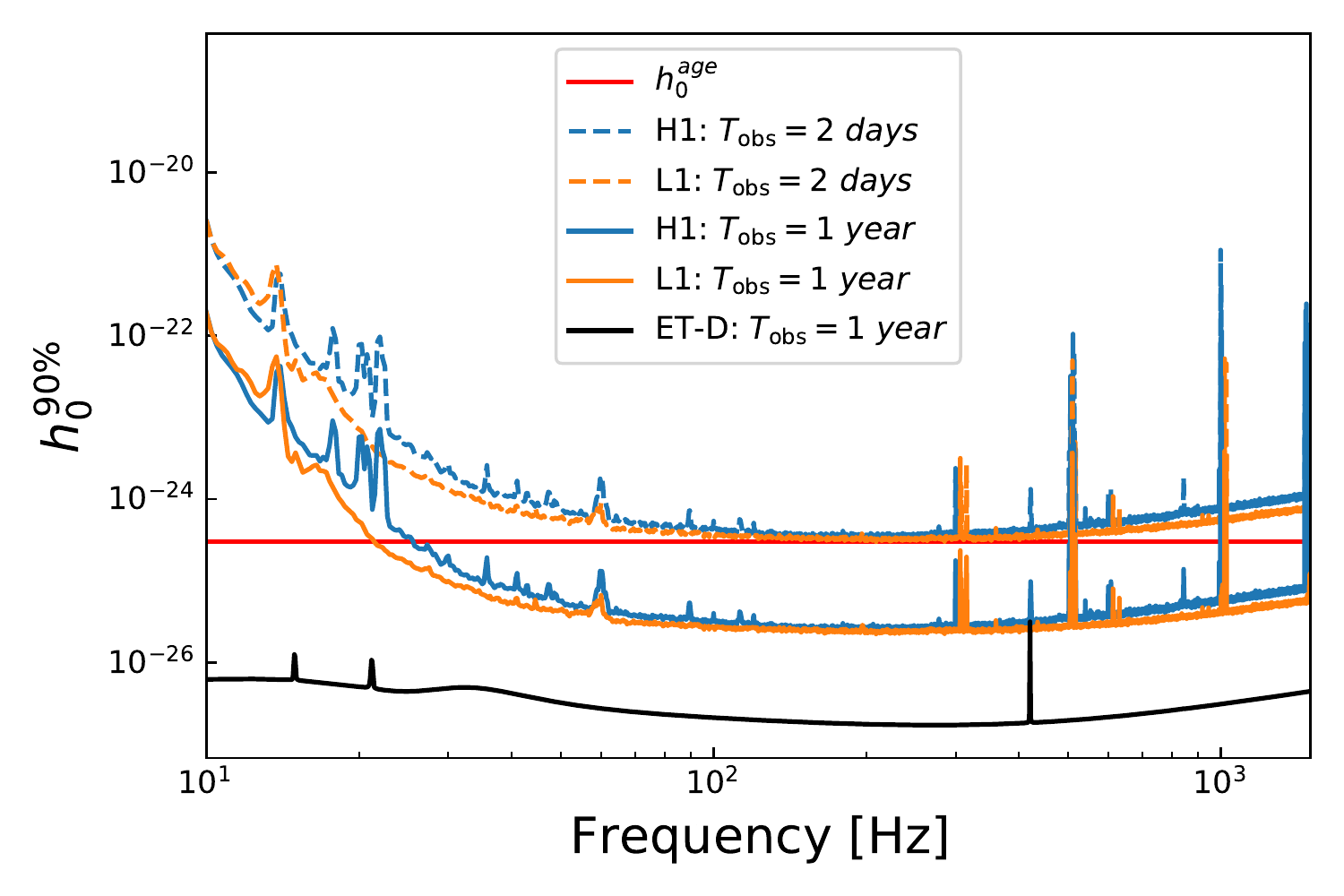}
    \caption{\label{fig:UL_th}
        Detectability of continuous gravitational waves from G4.8+6.2 with the current and future gravitational wave detectors. 
        The horizontal red line shows the age-based upper limit $h_{0}^{\text {age}}$. 
        Dashed and solid lines represent a different integration time of $95 \%$ confidence upper limits for Advanced LIGO (L1, H1) and the Einstein Telescope in configuration D (ET-D).
    }
\end{figure}

With a reasonable amount of time $T_{\mathrm{obs}}$ as shown in Fig.~\ref{fig:UL_th}, we can achieve a sensitivity at which it is theoretically possible to be detected.
In Fig.~\ref{fig:UL_th}, we plot the Advanced LIGO detectors sensitivity from the first three months of the O3 observation run \cite{LIGOnoise} and the future Einstein Telescope in the D configuration (ET-D) \cite{ET_D}.
We do not make use of Advanced Virgo data, as it is not expected to achieve a sensitivity comparable to Advanced LIGO during O3.

\section{CW Search Results}
\label{sec:Results}

In this section, we try to search the CW signals, while only upper limits are obtained.
Sections~\ref{sec:dsets},~\ref{sec:Fstat}, and~\ref{sec:param} describe the dataset, $\mathcal{F}$-statistic, and parameter spaces used in our search, respectively.
Sections~\ref{sec:Outliers} presents detection candidates (outliers) found in our search.
Several LALApps applications are from the open-source software package LALSuite \cite{lalsuite}.
Finally, we present the upper limits obtained in this search and comparison with other all-sky search pipelines in Sec.~\ref{sec:UL}.

\subsection{Datasets used}
\label{sec:dsets}

Computational costs restrict us to searching a limited time span $T_{\text {span}}$.
We choose $T_{\text {span}}$ without observation gaps for both the L1 and H1 detectors from the third observational run (O3b) that started on 2020-02-28 22:45:14 UTC (GPS time: 1266965132) and ended on 2020-03-01 23:00:14 UTC (GPS time: 1267138832).

We downloaded two days contiguous data using the distributed filesystem CernVM-FS \cite{CernVM}.
This filesystem allows us to mount GWOSC data locally on the user's computer.
We then used the code \texttt{lalapps\_Makefakedata\_v5} from LALSuite with options \texttt{--inFrames} and \texttt{--inFrChannels} to generate 1800-seconds SFTs with $50 \%$ overlap between each SFT.
All SFTs are Tukey-windowed with $\beta_{\text {Tukev}}=0.5$.

\subsection{$\mathcal{F}$-statistic}
\label{sec:Fstat}

The strain $\boldsymbol{x}(t)$ measured by a detector can be written as 
\begin{equation}
    \boldsymbol{x}(t)=\boldsymbol{n}(t)+\boldsymbol{h}(t ; \mathcal{A}, \lambda)
\end{equation}
where $\boldsymbol{n}(t)$ is the detector noise, and $\boldsymbol{h}(t ; \mathcal{A}, \lambda)$ is the CW signal.
The parameters of $\boldsymbol{h}(t ; \mathcal{A}, \lambda)$ can be split into two sets: the amplitude parameters $\mathcal{A}=\{h_{0}, \psi, \iota, \Phi_{0}\}$ and the phase-evolution parameters $\lambda=\{f, f^{(k)}, \alpha, \delta\}$, which are the intrinsic amplitude $h_{0}$, the wave polarization angle $\psi$, the inclination angle $\iota$, the initial phase $\Phi_{0}$, the wave frequency $f$, the spin down parameters $f^{(k)}$, and the sky position described by right ascension $\alpha$ and declination $\delta$ \cite{Whelan2014}.

CW signals with amplitude of order $10^{-25}$ will be buried by the noise.
In order to extract these signals, a matched filtering technique called the $\mathcal{F}$-statistic is used \cite{F1,F2,F3,F4,F5}.
The $\mathcal{F}$-statistic was derived as a maximum-likelihood estimator with respect to $\mathcal{A}$ \cite{prix2015}:
\begin{equation}
    \label{equ:likelihood}
    \mathcal{F}(\boldsymbol{x} ; \lambda) \equiv \max _{\mathcal{A}} \log \mathcal{L}(\boldsymbol{x} ; \mathcal{A}, \lambda)
\end{equation}
where $\mathcal{L}(\boldsymbol{x} ; \mathcal{A}, \lambda)=\frac{P\left(\boldsymbol{x} \mid \mathcal{H}_{S}(\mathcal{A}, \lambda)\right)}{P\left(\boldsymbol{x} \mid \mathcal{H}_{N}\right)}$ is the likelihood ratio comparing the signal hypothesis $\mathcal{H}_{S}(\mathcal{A}, \lambda)$ to a Gaussian-noise hypothesis $\mathcal{H}_{N}$.
Note that Eq.~(\ref{equ:likelihood}) can be computed analytically, but it still needs to maximize with respect to $\lambda$ to gain insight whether a signal with parameters $\mathcal{A}, \lambda$ is present in the data or not.
This can be determined by computationally expensive grid sampling covering the parameter space $\lambda$ with templates.
These templates must be placed densely enough so that for any possible signal, no more than a certain fraction of $\mathcal{F}$-statistic value is lost.

\subsection{Parameter space}
\label{sec:param}

Since the GW frequency of G4.8+6.2 is unknown, we searched a band of frequencies from $20$ to $1500 \mathrm{~Hz}$.
We used the characteristic age $\tau$ to estimate the largest magnitude of spin down (time derivative of the frequency) by $|\dot{f}| < f / \tau$.
Due to the short time spans, we neglected second and higher order terms and assumed that there were no glitches during this time.
We placed a grid in $(f, \dot{f})$ plane with fixed spacings, and the required grid spacings depend on $T_{\text {span}}$, as $1 / T_{\text {span}}$ and $1 / T_{\text {span}}^{2}$, respectively.
Given the age and distance of G4.8+6.2, NS would have moved only a few arcseconds from the geometrical center of the shell-type remnant even at transverse kick velocities of order $1000 \mathrm{~km} / \mathrm{s}$, while the sky resolution of the $\mathcal{F}$-statistic for the highest frequencies is much coarser than that (about an arcminute) \cite{Aasi2015,SNR15}.
% much coarser than the measured uncertainties in $\alpha$ and $\delta$, 
% The angular spacing of the sky grid points for the highest frequencies
Hence, we only searched a single sky position roughly at the center of G4.8+6.2.
The search parameters are summarized in Table~\ref{table:param}.

\begin{table}[h!]
    \centering
    \caption{\label{table:param} The key parameters used in our search of CWs from G4.8+6.2. Note that the spin down range depends on the frequency and on the characteristic age used.}
    \begin{tabular}{ll}
        \hline \hline Data span & $T_{\text {span}}=173700 \mathrm{~s}$ (O3b) \\
        Detectors & Hanford (H1) + Livingston (L1) \\
        Sky position & $\alpha=17^{\mathrm{h}} 33^{\mathrm{m}} 24^{\mathrm{s}}, \delta=-21^{\mathrm{d}} 34^{\mathrm{m}}$ \\
        Characteristic age $\tau$ & $860 \mathrm{~yrs}$ \\
        Frequency band & $f \in[20,1500] \mathrm{~Hz}$ \\
        Spin down range & $-f / \tau \leq \dot{f} \leq 0 \mathrm{~Hz} / \mathrm{s}$ \\
        Frequency resolution $\delta f$ & $5.76 \times 10^{-6} \mathrm{~Hz}$ \\
        Spin down resolution $\delta \dot{f}$ & $3.31 \times 10^{-11} \mathrm{~Hz} \mathrm{~s}^{-1}$ \\
        \hline \hline
    \end{tabular}
\end{table}

\subsection{Outliers}
\label{sec:Outliers}

\begin{figure}[h]
    \centering
    \includegraphics[width=0.5\textwidth]{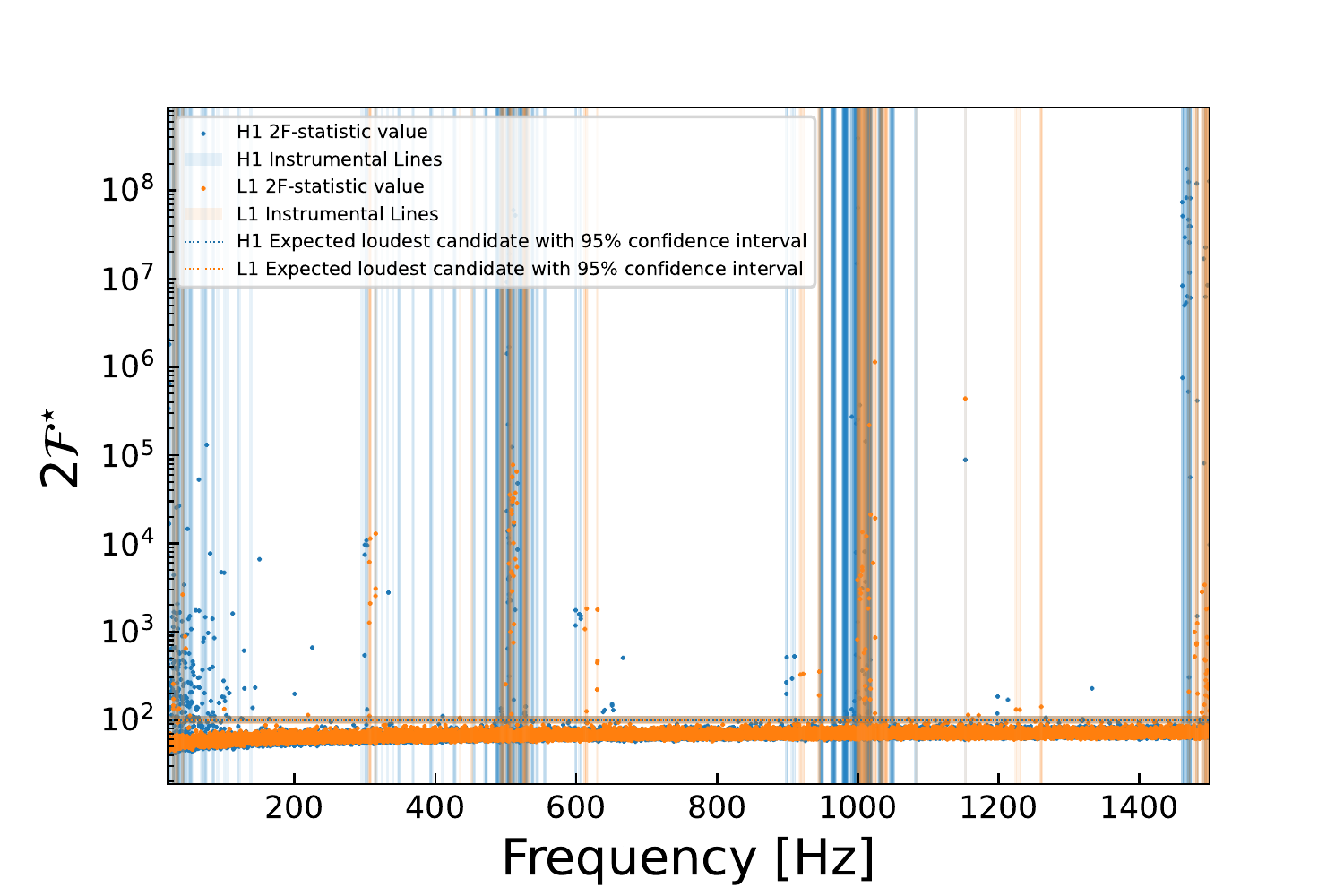}
    \caption{\label{fig:2F} Distribution of $2 \mathcal{F}^{\star}$ in each $0.1 \mathrm{~Hz}$ band for G4.8+6.2.
    The blue and orange indicate the Hanford and Livingston detectors, respectively.
    The frequency range is from $20$ to $1500 \mathrm{~Hz}$.
    The vertical stripes indicate known O3 instrumental lines. Dashed horizontal lines correspond to expected loudest candidate with $95 \%$ confidence interval.}
\end{figure}

We use \texttt{lalapps\_ComputeFstatistic\_v2} to compute the $\mathcal{F}$-statistic.
The search output is a list of the $2 \mathcal{F}$ statistic values for each parameter space point $\{f, \dot{f}\}$ [see Eq.~(\ref{equ:likelihood}) and Sec.~\ref{sec:param}].
Signals are expected to have a high value of $2 \mathcal{F}$. 
Therefore, we look for the loudest value in each $0.1 \mathrm{~Hz}$ frequency band and denoted it as $2 \mathcal{F}^{\star}$.
The distribution of $2 \mathcal{F}^{\star}$ in each subband is shown in Fig.~\ref{fig:2F}. 

\begin{figure}[h]
    \centering
    \includegraphics[width=0.5\textwidth]{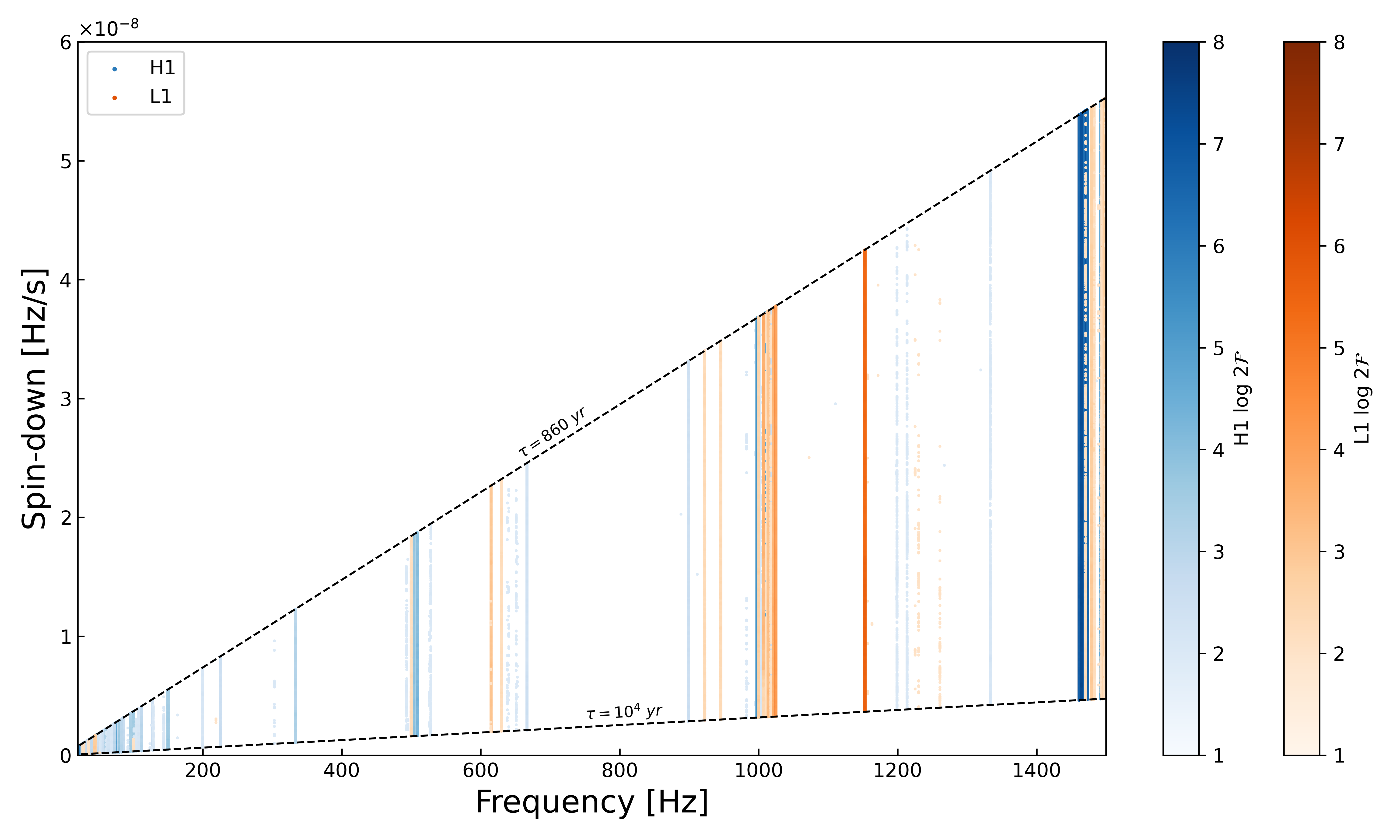}
    \caption{\label{fig:Outliers} Distribution of outliers in the frequency and spin down plane. The blue and orange indicate the Hanford and Livingston detectors, respectively. The color scale represents $\log 2 \mathcal{F}$. The dashed lines are the lines of constant characteristic age.}
\end{figure}

Data from the LIGO detectors are known to contain instrumental lines and glitches and may result in spuriously large values of the $\mathcal{F}$-statistic \cite{MilkyWay}.
To avoid the effects of these instrumental artifacts, we set a flat threshold on $2 \mathcal{F}$ at 98.6 for H1 and 98.0 for L1 and eliminate outliers overlapped with a list of known O3 instrumental lines \cite{O3Line}.
The threshold (i.e., the expected loudest $2 \mathcal{F}$ values in noise) was constructed based on fitting a Gumbel distribution to $2 \mathcal{F}^{\star}$ using \texttt{distromax} \cite{distromax}, where the frequency band was chosen from 660 to 880 Hz, which is in a spectral area that does not contain known instrumental lines.
All outliers with $2 \mathcal{F}$ values above the threshold and without overlapping with known instrumental lines are plotted in Fig.~\ref{fig:Outliers}.

We would expect that an astrophysical signal would be observed in both interferometers, and the observed $f/\dot{f}$ approximately equals the characteristic age $\tau$, at least within an order of magnitude.
With this filtering, no outliers survived.
Moreover, outliers scattered over the whole frequency band, which is inconsistent with the contours of an astrophysical signal \cite{Prix2007}.

\subsection{Upper limits}
\label{sec:UL}

\begin{figure}[h]
    \centering
    \includegraphics[width=0.45\textwidth]{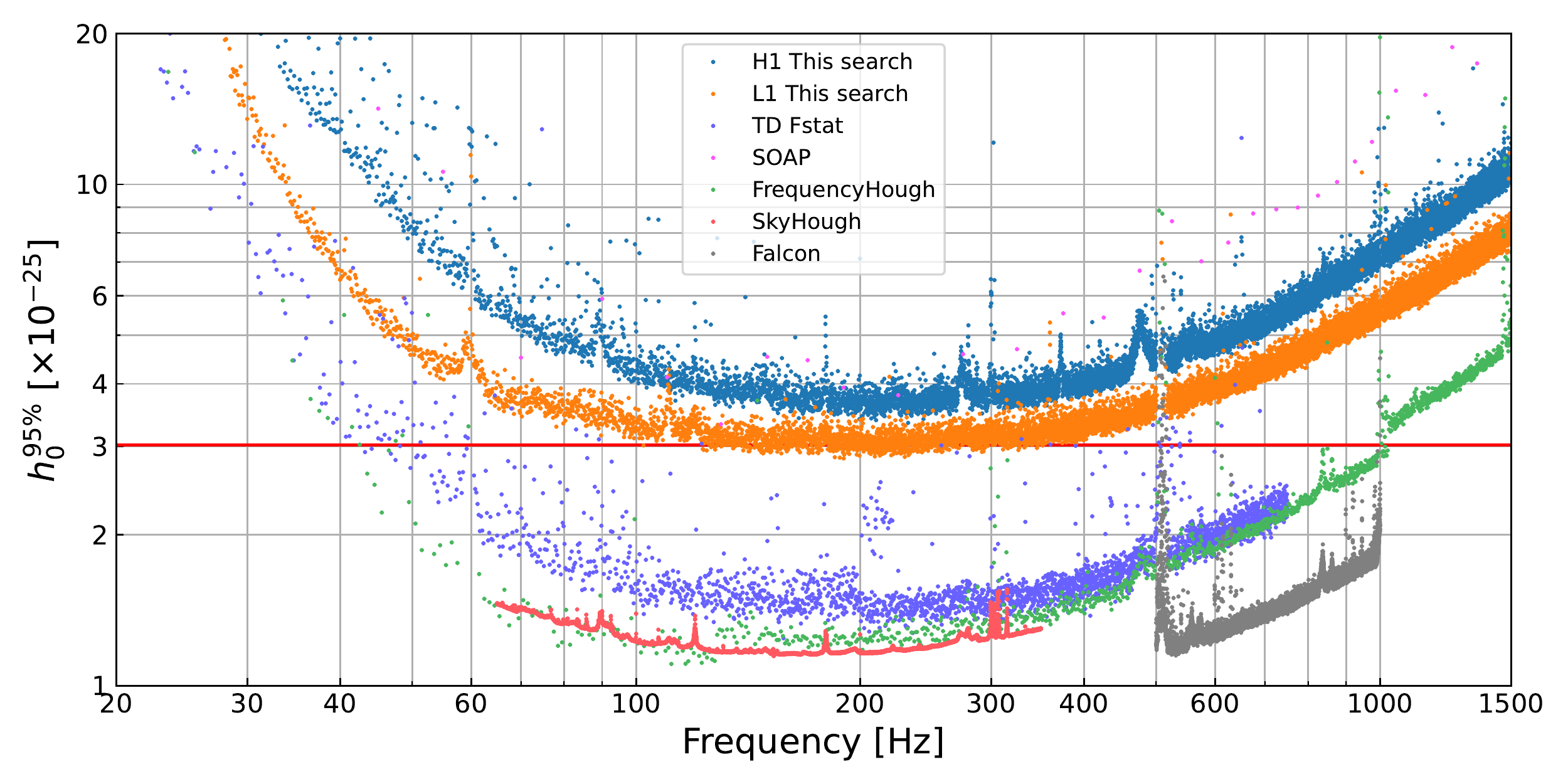}
    \caption{\label{fig:UL}
        The $95 \%$ confidence upper limit on signal strain $h_{0}^{95 \%}$ obtained in our search, together with those of previous all-sky searches.
        The different colors correspond to different pipelines.
        % Points represent the $95 \%$ confidence upper limit on signal strain $h_{0}^{95 \%}$ in each $0.1 \mathrm{~Hz}$ bands for G4.8+6.2.
        The horizontal red line shows the age-based upper limit $h_{0}^{\text {age}}$. 
        Note that the four pipelines FrequencyHough, SkyHough, Time-Domain F-statistic, and SOAP are population-averaged upper limits.
        Falcon and our results are in a fixed sky position roughly at the center of G4.8+6.2 (Data can be found in the Supplemental Material \cite{me}). 
    }
\end{figure}

There are no significant CW signals found in our search. 
So a $95 \%$ confidence level upper limit on signal strain has been computed over the full frequency band using \texttt{lalapps\_ComputeFstatMCUpperLimit}.
It performs software injections of simulated signals into the same datasets as used for the original searches and determines the required scale of $h_{0}$ at which $95 \%$ of signals would lead to such a value of $2 \mathcal{F}^{\star}$.
The results are plotted in Fig.~\ref{fig:UL}.
Our search barely beat the age-based upper limit, and a small number of $0.1 \mathrm{~Hz}$ bands failed to converge to an estimate for $h_{0}^{95 \%}$.

For comparison, Fig.~\ref{fig:UL} also shows upper limits from other five search pipelines: four from all-sky search pipelines using the LIGO-Virgo O3 observing run \cite{AllskyO3} and one from a frequency resolved atlas of the sky produced with the Falcon (Fast Loosely Coherent) pipeline \cite{Dergachev2019,Dergachev2022}.
The atlas data is stored in a computer readable format, which contains upper limits on the gravitational wave intrinsic strain $h_{0}$ as a function of signal frequency and source sky-position.
% We extracted the data set in the direction of the center of G4.8+6.2.
We extracted the data set with the spatial index (ra, dec)=(4.60367059707642, -0.376471847295761).
The Falcon search provides better upper limits on GW strain, but the signal-to-noise ratio of all points did not exceed the threshold set in the first stage of the Falcon pipeline.
It is worth noting that there are 1210 points in the second stage of the Falcon pipeline within 1 arcmin of our chosen spatial index, and 111822 points within 9 arcmin.
The downside is that Falcon only covers a frequency band of $500-1000 \mathrm{~Hz}$ with a frequency derivative up to $\pm 5 \times 10^{-11} \mathrm{~Hz} / \mathrm{s}$.
However, for a young magnetar, its frequency derivative is likely greater than the maximum value of $|\dot{f}|_{\max}$ used in Falcon \cite{Johnston2017}.

\begin{figure}[h]
    \centering
    \includegraphics[width=0.45\textwidth]{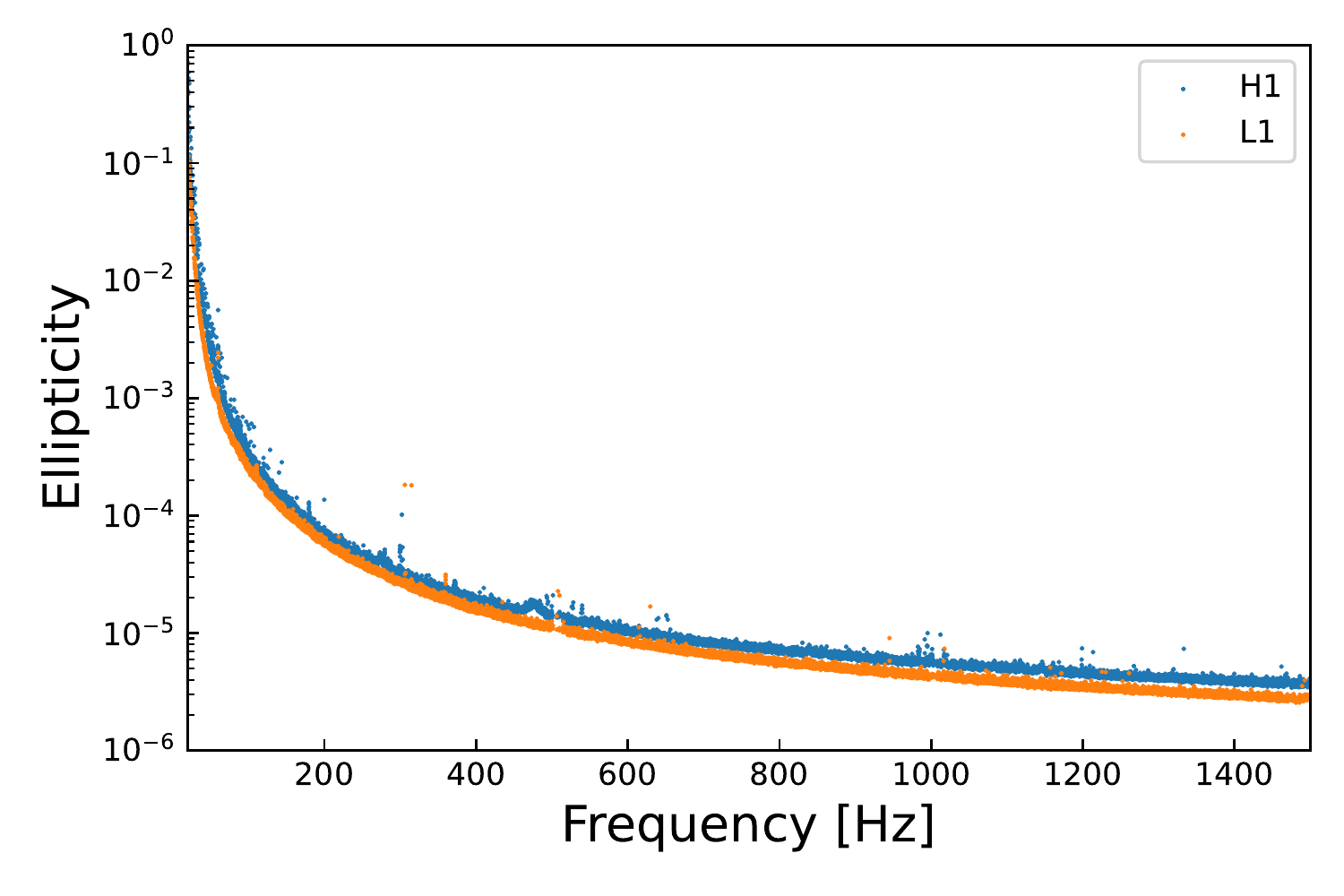}
    \caption{\label{fig:ellipticity}
        Upper limits on the ellipticity of the neutron star, derived from Eq.~(\ref{equ:ellipticity}) assuming a source distance of $8.1 \mathrm{~kpc}$ for G4.8+6.2.
    }
\end{figure}

Upper limits on $h_{0}$ can be converted to upper limits on the ellipticity of NSs $\epsilon=\left|I_{x x}-I_{y y}\right| / I_{z z}$  using
\begin{equation}
    \label{equ:ellipticity}
    \begin{aligned}
        \epsilon &=\frac{c^{4}}{4 \pi^{2} G} \frac{h_{0} D}{I_{z z} f^{2}} \\
        &\approx 7.66 \times 10^{-5} \left(\frac{h_{0}}{10^{-25}}\right)\left(\frac{100 \mathrm{~Hz}}{f}\right)^{2}\left(\frac{D}{8.1 \mathrm{~kpc}}\right)\left(\frac{10^{45} \mathrm{~g} \mathrm{~cm}^{2}}{I_{z z}}\right)
    \end{aligned}
\end{equation}
Unfortunately, our search results are not sufficient to put any meaningful constraints on the ellipticity as shown in Fig.~\ref{fig:ellipticity}, unless the source distance is closer than the estimated ${8.1} \mathrm{~kpc}$.

\section{Discussion and Conclusion}
\label{sec:Conclusion}

The isolated neutron star inside the center of the possible kilonova remnant G4.8+6.2 is a promising search target for continuous gravitational waves.
We found that the predicted age-based upper limits of G4.8+6.2 could be reachable using two days of Advanced LIGO O3b data.
Hence, we performed a search of continuous gravitational waves from the centers of G4.8+6.2 using the LALSuite software package.
The search covers a range of frequencies from $20$ to $1500 \mathrm{~Hz}$ and first frequency derivatives magnitude up to $10^{-8} \mathrm{~Hz} / \mathrm{s}$.
Although many outliers were found, the next filtering steps failed to confirm an astrophysical signal.
Hence, the primary results from this search are the upper limits on strain presented in Sec.~\ref{sec:UL}.
We also added several recent all-sky searches and the Falcon search.
The Falcon search provides better upper limits on GW strain; however, it is restricted in spin down range ($|\dot{f}|<5 \times 10^{-11} \mathrm{~Hz} / \mathrm{s}$).

Continuous gravitational waves have not yet been detected by any of search pipelines \cite{Piccinni2022}.
Consequently, we do not intend to do further CW searches using longer coherently integrated observation times due to the high computational cost.
In this work, we mainly illustrate the potential value of G4.8+6.2.
Detection of such signals would allow us to infer neutron star properties, such as moment of inertia, equatorial ellipticity, and the component of the magnetic dipole moment perpendicular to its rotational axis \cite{Lu2022}.
For massive millisecond magnetars, it prefers the merger scenario, and then probes the association between the NS, the KNR G4.8+6.2, and the guest star AD 1163.
For the current stage, it is still too early to see whether a BH or a NS is harbored inside the KNR. 
With upgrading of the GW detectors, we may first get even stringent upper limits, and constrain the eccentricity of NSs. 
Further, we may finally determine whether there is a NS or a BH.

It is also very promising to search for the neutron star within the center of G4.8+6.2 by multiwavelength electromagnetic observations.
The upcoming Chinese Space Station Telescope (CSST), a 2m space telescope in the same orbit as the China Manned Space Station, which is planned to be launched around 2024 \cite{Zhan2011}, could be able to make stringent upper limit on the optical KNR of G4.8+6.2.
The nondetection of a pulsar from G4.8+6.2 may come from the fact that the beaming of the radio emission is not pointing to the Earth. 
Considering if a young pulsar may produce repeating fast radio bursts, the sidelobe of the intense radio emission may reach the Earth. 
The 500-meter Aperture Spherical Radio Telescope \cite{Nan2011} may be able to detect such weak signal. 
However, the single dish may not be able to distinguish it from the nearby source, such as the Kepler SNR. 
Future Square Kilometre Array (SKA) \cite{Dewdney2009} may be able to resolve the region and find the radio emission from a single source, as well as provide further high signal-to-noise ratio observation on the radio emission of the remnant.

\begin{acknowledgments}
We thank the anonymous referee for the valuable comments.
We thank the helpful discussion with Ling Sun, Weihua Lei, Kai Wang, Enping Zhou, and Xilong Fan. 
This work is in part supported by the National Natural Science Foundation of China (Grant Nos. 12041306 and U1931203), by the National Key R\&D Program of China (2021YFC2203100), by MOE Key Laboratory of TianQin Project, Sun Yat-sen University. We also acknowledge the science research grants from the China Manned Space Project with No. CMS-CSST-2021-B11. 
The computation is completed in the HPC Platform of Huazhong University of Science and Technology, which costed about thousands of CPU core hours.
\end{acknowledgments}

% \nocite{*}
\bibliography{ref}
\bibliographystyle{apsrev4-1}

\end{document}